\documentclass[review, 12pt]{article}

\usepackage{graphicx,amsmath,amssymb,url,enumerate,mathrsfs,epsfig,color}
\usepackage{tikz}

\usepackage{pgfplots}
\usepackage{stmaryrd}
\usetikzlibrary{calc, patterns,arrows, shapes.geometric}
\pgfplotsset{compat=1.8}
\pgfarrowsdeclarecombine{twotriang}{twotriang}%
{stealth'}{stealth'}{stealth'}{stealth'}

\def\QED{\hskip0.1em\hfill\null\ \null\nobreak\hfill
\kern3pt\lower1.8pt\vbox{\hrule\hbox
{\vrule\kern1pt\vbox{\kern1.7pt \hbox{$\scriptstyle
QED$}\kern0.2pt}\kern1pt\vrule}\hrule}}
\newtheorem{note}{Note}[section]

\newtheorem{thm}{Theorem}
\newtheorem{defn}{Definition}

\title{Reflections on the structural and constitutive approaches to the theory of defects}
\author{Marcelo Epstein\footnote{University of Calgary, Calgary, Canada.}}
\date{}
\begin{document}
\maketitle
\begin{center}
To Yair Tene, fifty years later
\end{center}

\bigskip

{\bf Abstract}

An attempt is made to bring into harmony two of the  paradigms commonly used in the theory of continuous distributions of defects. It is shown that the common differential geometric apparatus is provided neatly by the theory of G-structures. In the case of a structural model, based on putative experimental observations at the microscopic level, a G-structure can be shown to emerge from the group of linear transformations that preserve a tensorial quantity. For the phenomenological (macroscopic) constitutive model, the G-structure arises from the notion of material isomorphism and the underlying local symmetry group of the constitutive law. A comparative example is presented in the framework of certain smectic liquid crystals.

\bigskip

{\bf Keywords}: Material isomorphism; uniformity; principal bundles; G-structures; liquid crystals; smectics.

\section{Introduction}

We will concern ourselves in these notes with two conceptually different paradigms that can be used to describe continuous distributions of defects in material bodies. These two paradigms can be roughly summarized as follows:

\begin{enumerate}
\item {\it The structural paradigm} is based on the imposition on the body of an additional geometric structure,  which, it may be argued, is observable under a putative microscope. This structure is represented usually by one or more tensor fields over the continuum. The standard case consists of three linearly independent vector fields, supposedly representing an observable atomic lattice. In non-strictly crystalline materials, such as certain smectic liquid crystals, one may postulate the existence of a 1-form, representing the partial arrangement of atomic planes typical of smectics. Some feature of this differential geometric structure is then used to define the presence of material defects.
\item {\it The constitutive approach}, steadfastly holding on to a purely macroscopic point of view, consistent with the epistemological view that a continuum theory does not recognize the existence of atomic structures per se, was promulgated by Noll \cite{noll} and Wang \cite{wang} in the 1960s. Assuming that all the information about the material constitution of a body should be entirely contained in the constitutive functionals observable at the macroscopic level, the concept of material isomorphism is used as the basis for the generation of a well-defined differential geometric structure, which may or may not be integrable, in a precise sense. The failure of integrability reveals the presence of defects.
\end{enumerate}

A third approach,  not included in our treatment here, is the {\it Volterra paradigm}, which considers the continuous limit of a sequence of Volterra cut-and-glue operations. It has been shown in \cite{raz} that this limit is, in a precise sense, equivalent to some version of the structural paradigm. It is noteworthy that the structural and constitutive paradigms, stemming from apparently disparate viewpoints, lead in many cases to identical criteria to detect the presence of continuous distributions of defects. This coincidence manifests itself in the fact that, although starting from different points of departure, they end up invoking the same geometrical apparatus.

When the structural paradigm is adopted, the absence of dislocations associated with the given tensor field (or fields) can be summarized as the answer to the following question: Is there a (local) chart such that the components of this tensor field (or fields) are constant throughout the chart? If the answer to this question is in the affirmative, we can reinterpret the chart as a local change of configuration and thereby obtain a tensor field invariant under Euclidean translations. If, for example, the tensor field is a differential 1-form, the condition for this to be the case is that the form must be closed. If, instead, the additional geometric structure consists of a field of bases, the condition is that the field must be holonomic, that is, the Lie bracket (or commutator) of each pair of the vector fields involved must vanish.

In contradistinction, the point of departure for the constitutive paradigm is the notion of {\it material isomorphism}. Roughly speaking, two material points are said to be materially isomorphic if their constitutive equations can be brought into exact correspondence by a mere distortion or, somewhat more precisely, by a change of local reference configuration. A body is {\it materially uniform} if all of its points are mutually materially isomorphic. A uniform body is {\it locally homogeneous} (or defect-free) if every point has a neighbourhood such that a change of configuration renders the constitutive law independent of position.

One of the obvious differences between these two paradigms is that the material symmetries encoded in the constitutive laws do not seem to play any role in the structural paradigm. To elucidate this and other aspects of the comparison of the two approaches, it is useful to carry out the formulation in the natural geometric setting provided by the linear frame bundle of the body manifold, and its associated bundles.

\section{The linear frame bundle and its associated bundles}

\subsection{Review of fibre bundles}

Recall that a material body $\mathcal B$ is defined as a smooth manifold of dimension $n=3$. As such, each material point $X \in {\mathcal B}$ is endowed with a tangent space $T_X{\mathcal B}$, which is an $n$-dimensional vector space. Each vector ${\bf v} \in T_X{\mathcal B}$ is an equivalence class of mutually tangent smooth curves through $X$. The formal union $T{\mathcal B}=\bigcup\limits_{X \in {\mathcal B}} T_X{\mathcal B}$ is itself a smooth manifold of dimension $2n$, which is called the {\it tangent bundle} of $\mathcal B$. We think of it as a smooth collection of tangent spaces, each of which is assigned to a point of $\mathcal B$. But it is precisely this assignation what gives a very special extra structure to the manifold $T{\mathcal B}$, and this is precisely what is meant by the word {\it bundle}. We think of $\mathcal B$ as the {\it base manifold} and of $T_X{\mathcal B}$ as the {\it fibre} over the point $X \in {\mathcal B}$

\begin{figure}
\begin{center}

\begin{tikzpicture}[scale=0.8]

\draw[ultra thick] (-0.75,0) -- (3,0);
\path [fill=lightgray] (0,0.3) to [out=90,in=-50] (-0.6,3.4) -- (0.5,3.4) to [out=-100,in=110] (0.75,0.3) -- (0,0.3);
\path [fill=gray] (0.75,0.3) to [out=110,in=-100] (0.5,3.4) -- (1.5,3.4) to [out=-70,in=110] (1.5,0.3) -- (0.75,0.3);
\path [fill=lightgray] (1.5,0.3) to [out=110,in=-70] (1.5,3.4) -- (2.6,3.4) to [out=-110,in=100] (2.25,0.3) -- (1.5,0.3);
\draw [thick] (0,0.3) to [out=90,in=-50] (-0.6,3.4);
\draw [thick] (0.75, 0.3) to [out=110,in=-100] (0.5,3.4);
\draw [thick] (1.5,0.3) to [out=110,in=-70] (1.5,3.4);
\draw [thick] (2.25,0.3) to [out=100,in=-110] (2.6,3.4);
\draw [thick] (3,0.3) to [out=80,in=-130] (3.5,3.4);
\draw [thick] (-0.75,0.3) to [out=90,in=-40] (-1.3,3.4);
\node at (3.0,-0.4) {$\mathcal B$};
\node at (0.75,-0.6){$\underbrace{\;\;\;\;\;\;\;{\mathcal U}_1 \;\;\;\;\;\;\;}$};
\node at (1.5,-1.2){$\underbrace{\;\;\;\;\;\;\;{\mathcal U}_2 \;\;\;\;\;\;\;}$};
\draw [thick] (0,-0.2) to [out=120,in=-120] (0,0.2);
\draw [thick] (0.75,-0.2) to [out=120,in=-120] (0.75,0.2);
\draw [thick] (1.5,-0.2) to [out=60,in=-60] (1.5,0.2);
\draw [thick] (2.25,-0.2) to [out=60,in=-60] (2.25,0.2);

\begin{scope}[shift={(6,0)}]

\draw[ultra thick] (-0.75,0) -- (3,0);
\path [fill=lightgray] (0,0.3) -- (0,3.4) -- (0.75,3.4) -- (0.75,0.3);
\path [fill=gray] (0.75,0.3) -- (0.75,3.4) -- (1.5,3.4) -- (1.5,0.3);
\path [fill=lightgray] (1.5,0.3) -- (1.5,3.4) -- (2.25,3.4) -- (2.25,0.3);
\draw [thick] (0,0.3) -- (0,3.4);
\draw [thick] (0.75, 0.3) -- (0.75,3.4);
\draw [thick] (1.5,0.3) -- (1.5,3.4);
\draw [thick] (2.25,0.3) -- (2.25,3.4);
\draw [thick] (3,0.3) -- (3,3.4);
\draw [thick] (-0.75,0.3) -- (-0.75,3.4);
\draw [ultra thick] (3.5,0.3) -- (3.5,3.4);
\node at (3.8,2.0) {$\mathcal F$};
\node at (3.0,-0.4) {$\mathcal B$};
\node at (0.75,-0.6){$\underbrace{\;\;\;\;\;\;\;{\mathcal U}_1 \;\;\;\;\;\;\;}$};
\node at (1.5,-1.2){$\underbrace{\;\;\;\;\;\;\;{\mathcal U}_2 \;\;\;\;\;\;\;}$};
\draw [thick] (0,-0.2) to [out=120,in=-120] (0,0.2);
\draw [thick] (0.75,-0.2) to [out=120,in=-120] (0.75,0.2);
\draw [thick] (1.5,-0.2) to [out=60,in=-60] (1.5,0.2);
\draw [thick] (2.25,-0.2) to [out=60,in=-60] (2.25,0.2);
\draw [step=0.75, black, thin] (-0.75,0.3) grid (3,3.4);
\node at (1.25,-0.18) {$X$};
\end{scope}
 \draw [-stealth'] (1.1,2.5) to [out=20, in=160] (7.3,3.2);
\draw [-stealth'] (1.1,2.5) to [out=5, in= 140] (7.3,1.8);
\draw [-stealth'] (7.3,3.2) to [out=-40, in= 40] (7.3,1.8);
\node at (4.5,3.7) {${ \psi}_{1}$};
\node at (4.5,3.0) {${ \psi}_{2}$};
\node at (8.,2.6) {${\psi}_{12}$};
\draw [-stealth'] (1.1,2.5) to [out=-110,in=110] (1.3,0.05);
\draw[-stealth'] (7.3,3.2)--(7.3,0.05);
\node[left] at (7.4,1.0) {$pr_1$};
\node at (1.25,-0.18) {$X$};
\node at (0.75, 1.3) {$\pi$};
%legend
\path [fill=lightgray] (4,-1) circle [radius=0.4];
\path [fill=lightgray] (4.4,-1) circle [radius=0.4];
\path[clip] (4,-1) circle [radius=0.4];
\fill[gray] (4.4,-1) circle [radius=0.4];

\end{tikzpicture}
\end{center}
\caption{Schematic representation of a bundle and two trivializations}
\label{fig:bundle}
\end{figure}
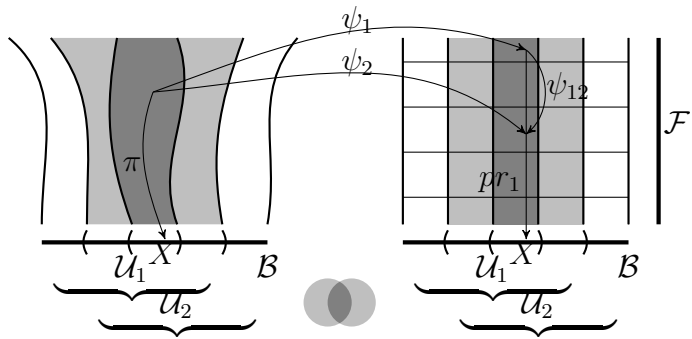

More generally, a smooth {\it fibre bundle} is a smooth manifold $\mathcal C$ and a smooth surjective map $\pi: {\mathcal C} \to {\mathcal B}$ onto a smooth manifold $\mathcal B$, called the base manifold. Technically, we require the projection to be a submersion. The inverse image ${\mathcal C}_X=\pi^{-1}(X)$ is called the fibre at $X\in {\mathcal B}$. We require, moreover, the condition of {\it local triviality}, which essentially requires that, locally, the fibre bundle resembles a Cartesian product. More specifically, for each point $X \in {\mathcal B}$, there is an open neighbourhood $\mathcal U \subset {\mathcal B}$ and a smooth and smoothly invertible map $\psi:\pi^{-1}({\mathcal U}) \to {\mathcal U} \times {\mathcal F}$, where $\mathcal F$ is a smooth manifold called the {\it typical fibre} of the bundle. This map $\psi$, called a {\it local trivialization}, is {\it fibre preserving}, namely, $pr_1\circ \psi = \pi$. In other words, if we apply the map $\psi$ to a point in the fibre over $X$ we obtain a point of the product that lies precisely over $X$, and not anywhere else. We have denoted by $pr_1$ the projection over the first factor of a Cartesian product. This definition sounds very intricate, but it is in fact very intuitive, as suggested in Figure \ref{fig:bundle}.

We are still not done with the definition, since we left the best for last. Noting that all the fibres ${\mathcal C}_X$ are mutually diffeomorphic (they all look the same, so to speak),\footnote{The diffeomorphism between fibres, however, is in general not canonical.} we ask what is the relation between two local trivializations whose domains have a non-empty intersection. If $\psi_1:\pi^{-1}({\mathcal U}_1) \to {\mathcal U}_1 \times {\mathcal F}$ and $\psi_2:\pi^{-1}({\mathcal U}_2) \to {\mathcal U}_2 \times {\mathcal F}$ are two local trivializations such that ${\mathcal U}_1 \cap {\mathcal U}_2 \ne \emptyset$, the {\it transition} between the first and the second is the map $\psi_{12}=\psi_2\circ \psi_1^{-1}: pr_1^{-1}({\mathcal U}_1 \cap {\mathcal U}_2) \to pr_1^{-1}({\mathcal U}_1 \cap {\mathcal U}_2)$. At each point $X$ of the intersection ${\mathcal U}_1 \cap {\mathcal U}_2$, the transition map is a transformation of $\mathcal F$. We will require that this transformation can only be construed as the left action of an element $g$ of a fixed group $\mathcal G$ acting smoothly on the left on $\mathcal F$, and that the dependence of $g$ on $X$ must be smooth. This group, which must be specified as part and parcel of the definition, is called the {\it structure group} of the bundle. In the case of the tangent bundle, the typical fibre is an $n$-dimensional vector space, and we allow all the possible linear automorphisms. Put differently, the typical fibre of the tangent bundle of a smooth $n$-dimensional manifold is ${\mathbb R}^n$, and the structure group is the general linear group $GL(n;{\mathbb R})$, which can be regarded as the group of all non-singular $n \times n$ matrices. The left action of the group on the typical fibre is, in this case, the matrix multiplication of a matrix times the column of the vector components in some basis. To summarize, a bundle consists of 5 ingredients and can be notated as $({\mathcal C}, {\mathcal B}, \pi, {\mathcal F}, {\mathcal G})$.

A {\it section} (or cross section) of a bundle  $({\mathcal C}, {\mathcal B}, \pi, {\mathcal F}, {\mathcal G})$. is a smooth map $\sigma: {\mathcal B} \to {\mathcal C}$
such that $\pi \circ \sigma = id_{\mathcal B}$. In words, a section assigns smoothly to each element $X$ of the base manifold an element of its fibre ${\mathcal C}_X$. In physical applications, a section is a {\it field} over $\mathcal B$. The nature of the field is given by the nature of the fibre. Thus, a section of the tangent bundle is a field of tangent vectors. A {\it local section} is a section of $\pi^{-1} ({\mathcal U})$, where $\mathcal U$ is a connected open subset of $\mathcal B$.

\subsection{Principal bundles}

Consider a bundle $({\mathcal C}, {\mathcal B}, \pi, {\mathcal F}, {\mathcal G})$, and let ${\mathcal F}'$ be manifold on which the group $\mathcal G$ has a smooth left action. Then, leaving everything else unchanged, including the local trivializations, we obtain a new fibre bundle $({\mathcal C}, {\mathcal B}, \pi, {\mathcal F}', {\mathcal G})$ with typical fibre ${\mathcal F}'$.\footnote{This construction is nicely explained and justified in \cite{steenrod}, p 43.} Two bundles related in this way are said to be mutually {\it associated}.

Since a group always has a natural left action on itself (by {\it left translation}), given a bundle $({\mathcal C}, {\mathcal B}, \pi, {\mathcal F}, {\mathcal G})$ we can always construct the associated bundle $({\mathcal C}, {\mathcal B}, \pi, {\mathcal G}, {\mathcal G})$, namely, a bundle whose typical fibre coincides with the structure group. This bundle is known as the {\it principal bundle} associated with the bundle of departure. It is essential, but not difficult, to show that in a principal bundle there is also a canonical {\it right action} of the structure group {\it on the total space} $\mathcal C$. In fact, the fibres can be regarded as the orbits of this action. The existence of this right action is traceable to the existence of two canonical actions of a group on itself, namely, the left and right translations. It is not difficult to show that if $p$ belongs to $\mathcal C$ the right action by a group element $g \in {\mathcal G}$ moves $p$ to a point $pg$ on the same fibre as $p$. If $\psi$ is a local trivialization, then we have that $pg=\psi^{-1}(\psi(p) g)$, where, with a clear abuse of notation, we use the apposition $ag$ to indicate the right action of $g$ on an element $a$, whether $a$ belongs to a fibre of $\mathcal C$ or to the typical fibre $\mathcal G$.

Starting from the tangent bundle $T{\mathcal B}$, consider at each point $X$ of $\mathcal B$ the collection $F_X{\mathcal B}$ of all bases of the tangent space $T_X{\mathcal B}$. The formal union  $F{\mathcal B}=\bigcup\limits_{X \in {\mathcal B}} F_X{\mathcal B}$ is a fibre bundle with the same trivialization maps as the tangent bundle. It is called the {\it (linear) frame bundle} of $\mathcal B$. A local section of this bundle is a {\it moving frame} (or {\it rep\`ere mobile}). At each point $X$ of an open set ${\mathcal U} \subset {\mathcal B}$, we have $n$ linearly independent tangent vectors ${\bf e}_1,...,{\bf e}_n$. Under a trivialization $\psi$, each of these vectors is mapped into a vector $\psi({\bf e}_\alpha)= A_{\;\;\alpha}^{\beta}\; {\bf g}_\beta$, where ${\bf g}_\beta\;(\beta=1,...,n)$ is the standard basis of ${\mathbb R}^n$, and where the summation convention is used. The components $A_{\;\;\alpha}^{\beta}$ form a non-singular matrix, which shows that the fibres are in a one-to-one correspondence with the general linear group $GL(n;{\mathbb R})$.\footnote{More technically, each fibre is a {\it$\mathcal G$- torsor}.} The transition functions between two trivializations amount to a left translation of the typical fibre $GL(n;{\mathbb R})$, as can be verified directly.

We conclude that the linear frame bundle is, in fact, the principal bundle associated with the tangent bundle. The right action over the bundle can be exposed by noticing that, having chosen arbitrarily a basis ${\bf e}_\alpha\;(\alpha=1,...,n)$ of $T_X{\mathcal B}$, each basis ${\bf f}_\beta$ at $X$ can be expressed as ${\bf f}_\beta=f^\alpha_{\;\;\beta}{\bf e}_\alpha$. The right action by an element $g^\mu_{\;\;\nu} \in GL(n;{\mathbb R})$ is the basis $R_g({\bf f})$ whose elements are $g^\mu_{\;\;\nu} {\bf f}_\mu=g^\mu_{\;\;\nu}\left( f^\alpha_{\;\;\mu} {\bf e}_\alpha\right)=\left(f^\alpha_{\;\;\mu} \,g^\mu_{\;\;\nu}\right) {\bf e}_\alpha$.

Other (non-principal) bundles associated with the frame bundle $F{\mathcal B}$ are the various tensor bundles. For example, the {\it cotangent bundle} $T^*{\mathcal B}$ is obtained by assigning to each point $X \in {\mathcal B}$ the dual space $T^*_X{\mathcal B}$ of $ T_X{\mathcal B}$. Higher order tensor bundles are obtained by respective tensor products of copies of $T{\mathcal B}$ and $T^*{\mathcal B}$.

\subsection{G-structures}

An important question in the theory of principal bundles is the possibility of {\it reducing} a principal bundle to another principal bundle whose structure group is a subgroup of the original. More precisely, let ${\mathcal P}=({\mathcal C}, {\mathcal B}, \pi, {\mathcal G}, {\mathcal G})$ be a principal bundle and let ${\mathcal G}'$ be a proper subgroup of $\mathcal G$. Consider another principal bundle of the form ${\mathcal P}'=({\mathcal C}', {\mathcal B}, \pi, {\mathcal G}', {\mathcal G}')$, that is, a principal bundle over the same base manifold $\mathcal B$ but with the smaller structure group ${\mathcal G}' \subset {\mathcal G}$. We say that ${\mathcal P}'$ is a {\it reduction} of $\mathcal P$ if there exists a fibre-preserving differentiable map $f:{\mathcal P}' \to {\mathcal P}$ such that
\begin{equation} \label{eq1}
f(p g) = f(p) g\;\;\;\;\;\;\;\;\forall p \in {\mathcal C}',\;\;g \in {\mathcal G}'.
\end{equation}

A useful way to imagine a reduction is to think of the reduced bundle ${\mathcal P}'$ as having as its total space a submanifold ${\mathcal C}'$ of $\mathcal C$, with the property of being closed precisely under the action of the subgroup ${\mathcal G}'$. This means that, under the right action of $g \in {\mathcal G}$ every point $p$ of the sub-fibre ${\mathcal C}'_X\subset {\mathcal C}_X$ has the property of remaining in this sub-fibre if, and only if, $g \in {\mathcal G}'$.\footnote{A rigorous proof of this assertion can be found in \cite{sternberg}, pp 296, 310.} In particular, a principal bundle is reducible to the trivial subgroup if, and only if, it admits a global section. For a principal bundle this condition is equivalent to the existence of a global trivialization.

\begin{defn} A G-structure on an n-dimensional base manifold $\mathcal B$ is a reduction of the frame bundle $F{\mathcal B}$ to a subgroup $G$ of the general linear group $GL(n;{\mathbb R})$.
\end{defn}

\section{The material G-structure}

\subsection{Introduction}

Having reviewed some basic geometric concepts, we proceed now to establish the relevance of G-structures in the theory of continuous distributions of dislocations. We will follow a two-pronged approach. In this section, we will arrive at the concept of a so-called {\it material G-structure}, uniquely (modulo conjugation) determined by the constitutive equation of a uniform body. In other words, we will be following the constitutive paradigm. In the next section, quite independently, we will rather adopt the structural paradigm and we will show how it too leads naturally to the emergence of a corresponding G-structure, which, for lack of a better term, we will call the {\it geometric G-structure}.

\subsection{Material uniformity}

For definiteness, we restrict the treatment to elastic constitutive equations. Following the standard notation of continuum mechanics, if we adopt a reference configuration for the body $\mathcal B$ and adopt thereat a (Cartesian) coordinate system $X^I\;(I=1,2,3)$, a {\it deformation} maps the body into the translation space of ${\mathbb R}^3$, with (Cartesian) coordinates $x^i\; (i=1,2,3)$, according to 3 smooth functions $x^i=\chi^i(X^1, X^2, X^3)$ with strictly positive Jacobian determinant. The {\it deformation gradient} $\bf F$  at a point $X$ is given in coordinates by the Jacobian matrix, whose entries are the partial derivatives $F^i_{\;\;I}=\partial \chi^i/\partial X^I$. An elastic constitutive equation specifies the components $s^{ij}$ of the Cauchy stress tensor $\bf s$ as smooth functions of the deformation gradient and the body point, namely, ${\bf s}= {\bf s}({\bf F}, X)$.
\begin{defn} Two body points, $X_1, X_2 \in {\mathcal B}$ are {\it materially isomorphic} if there exists a linear isomorphism ${\bf P}_{12}$ between their respective tangent spaces in the reference configuration such that $ {\bf s}({\bf F}, X_2)= {\bf s}({\bf FP}_{12}, X_1)$, identically for all deformation gradients $\bf F$.
\end{defn}
\begin{note} {\rm
This definition is the mathematical expression of the fact that the two points are made of the same material. Indeed, the constitutive equation depends subtly also on the reference configuration chosen. A change of reference configuration (which clearly does not affect the intrinsic material properties) results in the multiplication of the deformation gradient to the right by the Jacobian matrix of this change of reference. A constitutive response, therefore, is not to be identified with a particular function, but rather with a whole orbit of functions under the right action of the general linear group on the matrix argument. Two points are materially isomorphic if, and only if, their constitutive equations lie in the same orbit.}
\end{note}
We observe that a {\it material symmetry} $\bf G$ at a point $X$ is nothing but a material automorphism, that is, $\bf G$ satisfies the identity $ {\bf s}({\bf F}, X)= {\bf s}({\bf FG}, X)$. Moreover, given a material isomorphism ${\bf P}_{12}$ between $X_1$ and $X_2$, both ${\bf P}_{12} {\bf G}_1$ and ${\bf G}_2{\bf P}_{12}$ are also material isomoprhisms, where ${\bf G}_1$ and ${\bf G}_2$ are, respectively, material symmetries at $X_1$ and $X_2$. It follows that the symmetry groups of two materially isomorphic points are mutually conjugate, the conjugation being achieved by any material isomorphism between the two points.\begin{defn} A body is said to be {\it materially uniform} if all its points are materially isomorphic to each other. The body is {\it smoothly uniform} if every point has a neighbourhood over which the material isomorphism can be chosen as smooth functions of the material point $X$.
\end{defn}
The two definitions listed above were originally given by Noll \cite{noll}.

\subsection{Construction of the material G-structure}

Since material isomorphism is clearly an equivalence relation, when we are given a smoothly uniform body we may choose any point, $X_0$ say, as a {\it material archetype} for the whole body. If its constitutive equation is ${\bar{\bf s}}({\bf F})= {\bf s}({\bf F}, X_0)$, we can describe the global constitutive equation as
\begin{equation} \label{eq2}
 {\bf s}({\bf F}, X)={\bar{\bf s}}({\bf FP}(X)),
 \end{equation}
where ${\bf P}(X)$ is a material isomorphism from $X_0$ to $X$. This field cannot in general be chosen smoothly over the whole body, but it certainly can be so chosen over each member of an open covering of $\mathcal B$, since the body was assumed to be smoothly uniform.

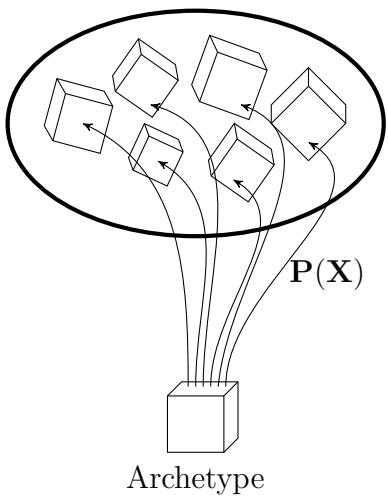
\begin{figure}
\begin{center}

\begin{tikzpicture}[scale = 0.5]
\begin{scope}[shift={(0,-3)}, scale=0.75]
\draw[black] (1,1) -- (1,-1) -- (-1,-1) -- (-1,1) -- (1,1); %square around the origin
\draw[black] (-1,1) -- (-0.5,1.5) -- (1.5,1.5) -- (1.5,-0.5) --
(1,-1); \draw[black] (1,1) -- (1.5,1.5);
\end{scope}
\begin{scope}[shift={(3,5)},rotate=45, scale=0.7]
\draw[black] (1,1) -- (1,-1) -- (-1,-1) -- (-1,1) -- (1,1); %square around the origin
\draw[black] (-1,1) -- (-0.5,1.5) -- (1.5,1.5) -- (1.5,-0.5) --
(1,-1); \draw[black] (1,1) -- (1.5,1.5);
\end{scope}
\begin{scope}[shift={(-3,5)},rotate=75,scale=0.65]
\draw[black] (1,1) -- (1,-1) -- (-1,-1) -- (-1,1) -- (1,1); %square around the origin
\draw[black] (-1,1) -- (-0.5,1.5) -- (1.5,1.5) -- (1.5,-0.5) --
(1,-1); \draw[black] (1,1) -- (1.5,1.5);
\end{scope}
\begin{scope}[shift={(1,6)},rotate=70,scale=0.7]
\draw[black] (1,1) -- (1,-1) -- (-1,-1) -- (-1,1) -- (1,1); %square around the origin
\draw[black] (-1,1) -- (-0.5,1.5) -- (1.5,1.5) -- (1.5,-0.5) --
(1,-1); \draw[black] (1,1) -- (1.5,1.5);
\end{scope}
\begin{scope}[shift={(-1,4)},rotate=65, scale=0.5]
\draw[black] (1,1) -- (1,-1) -- (-1,-1) -- (-1,1) -- (1,1); %square around the origin
\draw[black] (-1,1) -- (-0.5,1.5) -- (1.5,1.5) -- (1.5,-0.5) --
(1,-1); \draw[black] (1,1) -- (1.5,1.5);
\end{scope}
\begin{scope}[shift={(-1.3,6)},rotate=55, scale=0.6]
\draw[black] (1,1) -- (1,-1) -- (-1,-1) -- (-1,1) -- (1,1); %square around the origin
\draw[black] (-1,1) -- (-0.5,1.5) -- (1.5,1.5) -- (1.5,-0.5) --
(1,-1); \draw[black] (1,1) -- (1.5,1.5);
\end{scope}
\begin{scope}[shift={(1.25,3.75)},rotate=55, scale=0.6]
\draw[black] (1,1) -- (1,-1) -- (-1,-1) -- (-1,1) -- (1,1); %square around the origin
\draw[black] (-1,1) -- (-0.5,1.5) -- (1.5,1.5) -- (1.5,-0.5) --
(1,-1); \draw[black] (1,1) -- (1.5,1.5);
\end{scope}
\draw [ultra thick] (0,5) ellipse (5cm and 3cm); \draw[-stealth'] (0.4,-2) to
[out=90,in=-30] (1,3.5); \draw[-stealth'] (0.8,-2) to [out=90,in=-30]
(3,4.5); \draw[-stealth'] (0.0,-2) to [out=90,in=-30] (-1,4.0); \draw[-stealth']
(0.6,-2) to [out=85,in=-20] (1.2,5.5); \draw[-stealth'] (0.2,-2) to
[out=90,in=-15] (-1.2,5.5); \draw[-stealth'] (-0.2,-2) to [out=90,in=-25]
(-3.0,5.0);\node at(3.5,1.0) {${\bf P}({\bf X})$}; \node
at(0,-4.5) {Archetype};
\end{tikzpicture}
\end{center}
\caption{Material implants}
\label{fig:implants}
\end{figure}

If, as done in Figure \ref{fig:implants}, we place the archetype outside of the body, for convenience of the representation, we can visualize each material isomorphism ${\bf P}(X)$ as a {\it material implant} of the archetype. Just as in a surgical implant, the archetype is in general deformed to fit the damaged area, but the material properties are unaltered.

Consider the non-empty intersection of two open sets in the above mentioned cover. For each point of this intersection we have two different material isomorphisms with the archetype, say $\bf P$ and ${\bf P}'$, both from $X_0$ to $X$. But the composition ${\bf P}'{\bf P}^{-1}$ is a material isomorphism of the archetype with itself. Consequently, the point-wise transition functions over the non-empty intersection belong to the symmetry group of the archetype. We have thus constructed a principal bundle whose structure group is the symmetry group of the archetype. But recalling that material symmetry groups are subgroups of the general linear group, we have obtained a G-structure.

The presence of material defects (continuous distributions of dislocations, say) manifests itself by the lack of flatness of this G-structure, as presented, for example, in \cite{elzepsniat}. We will not go any further in this direction, since our main objective is only to compare the geometric structures entailed by the constitutive and structural paradigms. In this respect, an article by Mar\'in and de Le\'on \cite{mardel} presents many relevant details that are beyond this general presentation.

\section{The geometric G-structure}

\subsection{Tensors of type $H$}
\label{sec:tensors}

The traditional way of defining a tensorial quantity at a point of a manifold consisted of giving some indexed quantities (the components of the tensor in some coordinate system) and then stipulating how these components change upon a coordinate transformation. This supposedly old-fashioned definition can be embellished and dissimulated under the cloak of more modern terminology to produce the more general notion of a tensor field whose components at each point of an $n$-dimensional manifold $\mathcal B$ take values on an arbitrary vector space.

Let $H$ be a finite-dimensional real vector space on which the general linear group $GL(n;{\mathbb R})$ acts on the left. We start by forming the Cartesian product $F{\mathcal B} \times H$. To be sure, the elements of this product are the ordered pairs $(f, h)$, where $f \in F{\mathcal B}$ and $h \in H$. The first element $f$ is a basis of the tangent space at $\pi(f) \in {\mathcal B}$, while we can think of the second element as giving us the idea of components in some sense related to this basis. 

Inspired by the classical definition, we establish the following equivalence relation between pairs: $(f_1, h_1) \sim (f_2, h_2)$  if, and only if, there is a group element $g \in GL(n; {\mathbb R})$ such that $(f_2, h_2)= (f_1 g, g^{-1} h_1)$. We can recognize in this equivalence relation a kind of compensatory action, in the sense that a geometric object ($H$-tensor) manifests itself by means of components that depend on the basis chosen by counteracting the effect of a change of basis, which explains why the left action on $h$ is performed with the inverse of the group element that, through its right action on $f$, performs the change of basis. We can now consider the quotient space of this product by the equivalence relation, that is: ${\mathcal H}=F{\mathcal B} \times H/\sim$. The elements of $\mathcal H$ are precisely the equivalence classes themselves. It is not difficult to show that $\mathcal H$ is a bundle over $\mathcal B$ with typical fibre $H$ and structure group $GL(n;{\mathbb R})$. It is, therefore, a (non-principal) associated bundle of the linear frame bundle $F{\mathcal B}$. 

The cross sections of $\mathcal H$ are tensor (fields) of type $H$. Alternatively, and equivalently, we may regard these as entities defined directly over $F{\mathcal B}$, without reference to the associated bundle. In this case, we talk about a {\it tensor of type $H$ over $T{\mathcal B}$}. Such a tensor is a smooth map $t:F{\mathcal B} \to H$ satisfynig the identity
\begin{equation} \label{eq3}
t(fg)= g^{-1} t(f),\;\;\;\;\;\;\forall g \in GL(n;{\mathbb R})\;\;f \in F{\mathcal B}.
\end{equation}
We have worked on the principal frame bundle, but the same idea can be extended, mutatis mutandis, to any principal bundle.

\subsection{G-structures generated by tensors}
\label{sec:generated}

Although implicit in other treatments of G-structures, the notion of G-structures defined by tensors is explicitly introduced and treated in detail by Fujimoto \cite{fujimoto}. A somewhat broader characterization is given in \cite {mardel}.

Suppose that we arbitrarily fix an element $u$ of the vector space $H$. Since $GL(n; {\mathbb R})$ has been assumed to have a left action on $H$, we can define the {\it isotropy group} of $u$ as the set
\begin{equation} \label{eq4}
G_u=\{g \in GL(n; {\mathbb R})\;|\; gu=u \}.
\end{equation}
This is the largest closed subgroup of $GL(n;{\mathbb R})$ that leaves $u$ invariant. Consider now a $G_u$-structure ${\mathcal G}_u$, that is, a reduction of $F{\mathcal B}$ that happens to have $G_u$ as its structure group. Let us define the constant map $t: {\mathcal G}_u \to H$ given by $t(f)=u$ for all $f \in {\mathcal G}_u$. This map can be considered a tensor of type $H$ defined on ${\mathcal G}_u$. Indeed, all pairs $(f,u)$ over each fibre are related by the equivalence relation $\sim$, since the structure group has been chosen precisely to leave $u$ invariant.

We would like to characterize a G-structure by means of a vector field on $F{\mathcal B}$, but so far we have been going in the opposite direction. Nevertheless, to gain a better appreciation of what kind of properties a tensor over $F{\mathcal B}$ must have so as to determine a G-structure, let us extend the constant map $t$, defined so far on ${\mathcal G}_u$ only, to the whole frame bundle. By the basic property (\ref{eq3}) of tensors of type $H$, the extension $t:F{\mathcal B} \to H$ can only be given by
\begin{equation} \label{eq5}
t(f) = g^{-1} u,
\end{equation}
while $g$ runs over the whole general  linear group. Clearly, $t$ takes values only on the orbit $H_u$ of $u$, namely,
\begin{equation} \label{eq6}
H_u =\{gu\;|\;g \in GL(n;{\mathbb R})\}.
\end{equation}
We assume the map $t$ to be differentiable.\footnote{See \cite{fujimoto}, p 22, for a proof of this fact and of the following theorem.}

These two properties of the tensor $t$, namely, that it takes values on an orbit of $H_u$ and that it is differentiable, are enough to completely define a G-structure whose structure group is the isotropy group of $u$.
\begin{thm} Giving a G-structure on $\mathcal B$ is the same as giving a tensor $t:F{\mathcal B} \to H$ of type $H$ on $F \mathcal B$ which satisfies the following two conditions:
\begin{enumerate}
\item $t$ takes values on  $H_u$;
\item $t$ is a differentiable map (with an appropriate differentiable structure in $H_u$).
\end{enumerate}
\end{thm}
Given a tensor of type $H$ with this property, the corresponding $G_u$ structure is the inverse image of $u$, namely, ${\mathcal G}_u=t^{-1}(u)$. In practical applications, one may be able to choose a moving frame such that the components of the given tensor field are constant. This moving frame is in the $G_u$ structure.

\subsection{Constructing the geometric G-structure}

The basic tenet of the structural approach is the existence of an additional geometric object superimposed on the body manifold. This object is supposedly observable experimentally by microscopy. If we assume it to be tensorial, as one would expect in applications, we can generate the corresponding G-structure, as defined above, and as illustrated in the following example.

\section{An example}

\subsection{A structural description of a smectic A}

It is never easy to translate into mathematical terms a complex physical phenomenon and to distinguish the essential from the accessory. The case of smectic liquid crystals, however, provides an excellent example of clarity in this regard. Thus, we read in \cite{chandrasekhar}
\begin{quotation}
Smectic liquid crystals have stratified structures but a variety of 
molecular arrangements are possible within each stratification. In smectic 
A the molecules are upright in each layer with their centres irregularly 
spaced in a `liquid-like' fashion ... The interlayer attractions 
are weak as compared with the lateral forces between molecules and in 
consequence the layers are able to slide over one another relatively easily. 
\end{quotation}

Or, in added detail, as in \cite{kamien},

\begin{quotation}
Smectic liquid crystals consist of rod-shaped molecules that spontaneously form both directional (nematic) order and a one-dimensional density wave, commonly described as a layered system; the spacing between the layers is approximately the rod length, $a$. In two dimensions, we can picture the layers as a set of nearly equally spaced curves lying in the plane. The ground states are characterized by both equal spacing between these curves and vanishing curvature.
\end{quotation}

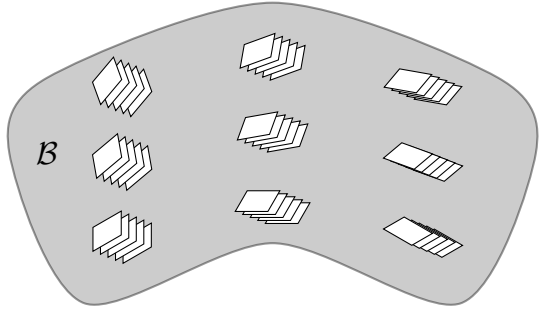
\begin{figure} [h]
\begin{center}
\begin{tikzpicture}
{\begin{scope}[ shift={(0.8,0.2)}]
\draw [thick, fill=gray, opacity=0.4] plot [smooth cycle] coordinates {(-0.5,0)  (-1.5,2)  (-1,3)  (2,4)  (5,3)  (5.5,2)  (4.5,0)  (2,0.8)};
\node at (-1.,2) {$\mathcal B$};
\end{scope}}
\foreach \y in {0,1,2}
\foreach \x in {0,1,2}
{\begin{scope}[rotate=0, shift={(2*\x,\y+0.4*\x*(2-\x))}]
\foreach \z in {0,1,2,3,4}
\node [draw, fill=white, rotate=30-30*\x+10*\y, trapezium left angle=60, trapezium right angle= 120, minimum height=0.5, minimum width= 0.5, trapezium] at (1-0.1*\z,1+0.04*\z) {};
\end{scope}}
\end{tikzpicture}
\end{center}
\caption{Pictorial representation of a differential 1-form on $\mathcal B$}
\label{fig:form}
\end{figure}

It is clear from these physical descriptions that smectics are not necessarily representable by Bravais lattices. In the continuous limit, we can conjecture that a smectic  is faithfully represented by a differential 1-form. Indeed, as suggested in another physical context \cite{misner}, if we picture a vector as an arrow, then a covector should be represented by a stack of parallel planes endowed with a certain density. The action of a covector on a vector can be visualized as the `number' of planes pierced by the arrow. Recall that a differential 1-form is a smooth field of covectors. Consequently, a differential 1-form $\omega$ can be understood as the smooth specification of a stack of planes at each point of $\mathcal B$, as shown schematically in Figure \ref{fig:form}. The resemblance to the physical descriptions of a smectic A is obvious.

Suppose that this layering 1-form $\omega$ has been constructed on the basis of smoothed-out averages of microscopic observations, without any prejudice to constitutive equations of any kind. As a section of the cotangent bundle $T^*{\mathcal B}$ (which is associated to the bundle of frames), we conclude that it gives rise to a tensor over $F{\mathcal B}$, in the sense explained in Section \ref{sec:tensors}. We can adopt a moving frame  ${\bf e}_1,{\bf e}_2,{\bf e}_3$ (that is, a basis of $T_X{\mathcal B}$ at each point $X$) such that its dual basis, $e^1,e^2,e^3$, has $\omega=e^1$ as its first element. In this system, therefore, the components of $\omega$ are given by the row vector $<\omega>=<1,0,0>$. The result of the left action of a matrix $[g] \in GL(3;{\mathbb R})$ on a row vector $<\omega>$ is the row vector $<\omega> [g]^{-1}$ (since the action on a column vector $\{v\}$ is $[g]\{v\}$, and the linear evaluation of covectors on vectors must be preserved). We conclude that the isotropy group $G_\omega$ of our $<\omega>=<1,0,0>$ (namely, the collection of matrices such that  $<\omega> [g]^{-1}=<\omega>$) consists of all the matrices $[k]$ of the form
\begin{equation} \label{eq7}
[k]=\left[
\begin{matrix}
1&0&0 \\ a&b&c\\d&e&f
\end{matrix}
\right]
\end{equation}
For physical reasons, we may want to further restrict this isotropy group by demanding that, once a co-frame $e^\alpha\;(\alpha=1,2,3)$ (that is, 3 linearly independent covectors) has been chosen, the volume form $e^1\wedge e^2 \wedge e^3$ must be preserved. The implication of this additional condition is that the determinant of the matrix $[k]$ must be equal to 1.

According to the notion of a G-structure defined by a tensor, as described in Section \ref{sec:generated}, we conclude that the geometric  G-structure generated by our differential form consists of the given basis ($e^1=\omega, e^2, e^3$) and all the bases obtained from it via the right action of the isotropy group just described.

\subsection{A constitutive model}

The constitutive response of a hyperelastic material can be completely characterized by a single scalar function $\psi$ of the deformation gradient $\bf F$. The principle of material frame indifference requires that this dependence be mediated exclusively by the (symmetric) right Cauchy-Green tensor $\bf C={\bf F}^T{\bf F}$. Clearly, for this definition to make sense, the Euclidean metric is presupposed. Be that as it may, in the constitutive approach we are at freedom to choose any function $\psi=\psi({\bf C})$ that can, in principle, be determined from macroscopic experiments in a laboratory, without any need to invoke microstructural considerations or microscopic observations.

Accordingly, we can imagine an experimental setup, perhaps in a biomaterials laboratory, capable of carrying out accurate measurements with soft materials. The sample would be aligned with a machine-related Cartesian coordinate system and then subjected to arbitrary values of the Cauchy-Green tensor with Cartesian components $C_{IJ}$ measured with respect to the unloaded reference configuration. Suppose that the execution of many experiments on the given sample leads to the following reasonably accurate formula for the free energy $\psi$ as a function of two scalar variables
\begin{equation} \label{eq8}
\psi=\psi(r, d),
\end{equation}
where $d=\det{\bf C}$, and where
\begin{equation} \label{eq9}
r=C_{22} C_{33}-C_{23}C_{32}=C_{22}C_{33}- C_{23}^2.
\end{equation}

An arbitrary function of these variables has precisely the material symmetries given by Equation (\ref{eq7}), with the additional condition that $\det [k]=1$. Therefore, the material G-structure of a uniform body modeled after the archetypal constitutive equation (\ref{eq8}) is precisely a reduction of the frame bundle $F{\mathcal B}$ to this subgroup of $GL(3;{\mathbb R})$. Consequently, the material G-structure coincides with the geometric G-structure.

\section{Final thoughts}

A shrewd inspection of the constitutive equation (\ref{eq8}) would reveal that the independent variable $r$ can be regarded as the square of the magnitude of the cross product of the deformed images of two initially orthonormal vectors aligned with the referential coordinate axes $X^2$ and $X^3$. It would appear, therefore, that there has been a certain amount of borrowing of information from the microscopic level. This may well have been the case, historically speaking. Nevertheless, we have already witnessed other historical instances when the process was actually reversed. Think, for example, of the atomic theory, not so much the one that grew spontaneously out of the unique genius of the ancient Greek mind, but that which arose out of observations at the macroscopic level of chemical experiments by John Dalton (1766-1844) and Amedeo Avogadro (1776-1856). Atoms had to be invented before they could be observed more directly, two or three generations after the pioneers. A similar historical phenomenon occurred when Max Planck (1858-1947) reluctantly posited the quantized nature of electromagnetic radiation.

Within the more mundane framework of these notes, we may venture to say that, if a constitutive equation such as (\ref{eq8}) had been available at the early stages of the discovery of liquid crystals, one would have been led to suspect that there are preferred planes of some sort at the microscopic level, as it  turned out to be the case for smectics.\footnote{The interrelation between scientific levels of discourse and its implications for the question of scientific explanation have been considered at length by the noted Canadian philosopher of science Mario Bunge \cite{bunge}. Bunge distinguishes between subsumptive (or phenomenological) and interpretive explanations, the latter involving a deeper analysis at more basic levels of reality than the level to which the explanation belongs. With scholastic irony, Bunge calls the former `explicatio obscurum per obscurius'.} What seems to be remarkable is that at least some of the detailed geometric information that is made available at the microscopic level  is somehow encapsulated in the macroscopic (phenomenological) constitutive equation. The theory of G-structures is the common differential geometric apparatus that harmonizes  the two paradigms with each other.

\end{document}